\documentclass[letterpaper,11pt]{article}
\pdfoutput=1
\usepackage{jheppub}
\usepackage{amssymb,amsmath,graphicx}

\newcommand{\w}{\bar{w}}
\newcommand{\E}{\mathcal{E}}
\newcommand{\h}{\mathcal{H}}
\newcommand{\ket}[1]{\left|#1\right>}
\newcommand{\bra}[1]{\left\langle#1\right|}

\title{Time dependence of reflected entropy in rational and holographic conformal field theories}
\author{Mudassir Moosa}%
\affiliation{Department of Physics, Cornell University, Ithaca, NY, 14853, USA}
\emailAdd{mudassir.moosa@cornell.edu}

\abstract{
We calculate the time dependence of the reflected entropy of two disconnected regions after a global quench in $(1+1)$-dimensional conformal field theories and in large temperature limit. For rational conformal field theories, we find that the time evolution of the reflected entropy is the same as that of the mutual information. We get the same result for holographic theories in the limit where the separation between disconnected regions is much smaller than their respective sizes. We discuss how this result is consistent with the quasi-particle picture of Calabrese and Cardy \cite{calabrese-cardy}.}
	
\begin{document}
\maketitle


\section{Introduction}
\label{intro}

Quantum quenches have proven to be an interesting tool to understand how non-equilibrium systems thermalize. In a quantum quench, we start with a ground state of some Hamiltonian $H_{0}$ and at time $t=0$ we change the Hamiltonian from $H_{0}$ to $H$. 
The state for $t > 0$ evolves according to the new Hamiltonian, $H$, and will have a non-trivial time dependence. 

An interesting example of a quantum quench is when 
$H_{0}$ is the Hamiltonian of some gapped theory whereas $H$ is the Hamiltonian of a CFT. In this case, it was argued in \cite{calabrese-cardy,gq-2} that we can model the quantum quench by replacing the state at $t=0$ by 
\begin{align}
    \ket{\Psi(t=0)} \, = \, e^{-\frac{\beta}{4}H} \, \ket{\mathcal{B}} \, ,\label{eq-bdy-quench}
\end{align}
where $\ket{\mathcal{B}}$ is a conformal boundary state and $1/\beta$ corresponds to the mass gap of the original theory. 
The state at time $t \ge 0$ is then given by
\begin{align}
    \ket{\Psi(t)} \, = \, e^{-\left(it + \frac{\beta}{4}\right)H} \, \ket{\mathcal{B}} \, .
\end{align}
Even though the state of the whole system remains pure (under unitary evolution), we expect the reduced state of some small subsystem to thermalize at late times. This is exactly what was observed in the behavior of the correlation functions in \cite{calabrese-cardy,gq-2}. Further evidence of (local) thermalization comes from the time evolution of the entanglement entropy of a subregion $A$ of size $L$, which is defined as 
\begin{align}
S_{A}(t) \, = \, - \, \text{tr} \, \rho_{A}(t) \, \log \rho_{A}(t) \, \quad\quad \text{where} \quad\quad \, \rho_{A}(t) \, = \, \text{tr}_{\bar{A}} \ket{\Psi(t)}\bra{\Psi(t)} \, .
\end{align}
In the scaling limit
\begin{align}
    t \, , L \, \gg \beta \, ,\label{eq-scaling}
\end{align}
it was found in \cite{calabrese-cardy,gq-2} that the time evolution of the entanglement entropy, for all CFTs, only depends on the
central charge, $c$, of the CFT and the parameter $\beta$ of the initial state. In particular, it was found that the entanglement entropy of region $A$ as a function of time is given by 
\begin{align}
    S_{A}(t) \, = \, S_{A}^{\text{vac}} \, + \, 2  s_{eq} \times \label{eq-sa}
\begin{cases}
\, t \quad&\text{for $t <\frac{L}{2}$} \, ,\\[0ex]
\, \frac{L}{2} \quad&\text{for $t>\frac{L}{2}$} \, ,
\end{cases}
\end{align}
where $S_{A}^{\text{vac}}$ is the vacuum entanglement entropy at $t=0$ which contains the usual ultraviolet divergence \cite{Holzhey:1994we}, and 
\begin{align}
    s_{eq} \, \equiv \,  \frac{\pi c}{3\beta}  \label{eq-seq}
\end{align}
is the thermal entropy density at temperature $1/\beta$. 

A simpler model for studying time-dependence and (local) thermalization, called the `thermal double model', was introduced in \cite{hartman-maldacena}. In this model, we take two copies of our CFT, \textit{i.e.} CFT$_{1} \, \otimes $ CFT$_{2} \, $,  and consider the following entangled state:
\begin{align}
 \ket{\Psi_{\beta}} \, = \, \frac{1}{\mathcal{N}_{\beta}} \, \sum_{n} \, e^{-\beta E_{n}/2} \, \ket{n}_{1}\otimes\ket{n^{*}}_{2} \, ,   \label{eq-tfd}
\end{align}
where $\ket{n}$ are the energy eigenstates of the original CFT, $\ket{n^{*}}$ are the action of the antiunitarity CPT on $\ket{n}$, and $E_{n}$ are the corresponding energy eigenvalues. Furthermore, we demand that the time evolution is generated by $H_{1} + H_{2}$. As a result, the state in Eq.~\eqref{eq-tfd} evolves in time. 

Now suppose that the subregion $A$ consists of two identical intervals of size $L$, one in each copy of the CFT. The time dependence of the entanglement entropy for region $A$ in this model was studied in \cite{hartman-maldacena}. It was found that this time dependence, up to a factor of $2$, is the same as the time dependence in Eq.~\eqref{eq-sa}. 

The quantitative behavior of $S_{A}(t)$ in these two models, that is the linear growth for $t < L/2$ and the saturation for $t>L/2$, can be described in terms of the propagation of entangled pairs of quasi-particles \cite{calabrese-cardy,gq-2}. Assume that EPR pairs of entangled quasi-particles are uniformly produced everywhere at $t=0$. Each quasi-particle and its entangled partner move in the opposite direction with (instantaneous) speed $v=1$. The entanglement entropy of region $A$ at any time $t$ is proportional to the number of EPR pairs for which one entangled partner is in region $A$ at time $t$ whereas other is outside the region $A$. 

Now consider two disconnected subregions, $A$ and $B$. The entanglement entropy for disconnected regions are not completely fixed by the conformal symmetry and hence depends on the details of the CFT \cite{gq-11,gq-12,hartman-2}. Nevertheless, it was shown in \cite{hartman-2} that the quasi-particle picture correctly captures the evolution of entanglement entropy for disconnected regions for a certain class of theories. In these theories, the asymptotic number of conserved currents is approximately equal to the total number of states. In other words, the central charge of these theories is $c \, = \, c_{\text{current}}$, where $c_{\text{current}}$ is an effective central charge of the chiral sector of the theory.
For this reason, these kind of theories were called `current dominated' in \cite{hartman-2}. Examples of current dominated theories include all rational CFTs and some non-rational CFTs \cite{hartman-2}.

Another class of CFTs that we would be interested in is holographic theories. These theories have $c \gg c_{\text{current}} \sim 1$ and hence, these CFTs are not current dominated. Indeed, the quasi-particle picture is known to be incorrect for these theories \cite{gq-11,gq-12,hartman-2}. The time dependence of entanglement entropy for these theories has been studied in \cite{gq-3,Albash:2010mv,hartman-2,gq-4,gq-5,gq-6,gq-7,gq-9,gq-10,gq-11,gq-12,gq-13,gq-14,gq-15,gq-16,gq-17,gq-18,gq-20,gq-22}. The quantitative behavior of the entanglement entropy in holographic theories can be described in terms of a spread of an `entanglement tsunami wave' \cite{gq-9,gq-10,gq-12} or in terms of a `minimal membrane' \cite{Mezei-1,Mezei-3}. 

It is also interesting to study how does the entanglement or correlation between two disconnected regions, $A$ and $B$, change following a quantum quench. Entanglement entropy $S_{A\cup B}(t)$ is not a useful quantity for this purpose. This is because $S_{A\cup B}(t)$ measures the entanglement of $A$ and $B$ with the rest of the system instead of measuring the entanglement between $A$ and $B$. One possible quantity that captures the correlation between two disconnected regions is the mutual information, which is defined as
\begin{align}
    I(A|B) \, \equiv \, S_{A} + S_{B} - S_{A\cup B} \, .
\end{align}
For current dominated theories, time evolution of the mutual information can also be described in terms of propagating quasi-particles \cite{10.21468/SciPostPhys.4.3.017}. In particular, the mutual information at any time is proportional to the number of EPR pairs for which one entangled partner is in region $A$ whereas the other is in region $B$. For concreteness, suppose that regions $A$ and $B$ are of equal size $L$ and they are separated by a distance $\ell$. The mutual information in the thermal double model, according to the quasi-particle picture, is \cite{10.21468/SciPostPhys.4.3.017}
\begin{align}
    I(A|B)(t) \, = \, 4 s_{eq} \, \times \label{eq-mi}
\begin{cases}
\, 0 \quad&\text{for $\quad t <\frac{\ell}{2}$} \, ,\\[0ex]
\, t - \frac{\ell}{2} \quad&\text{for $\quad \frac{\ell}{2} < t < \frac{L+\ell}{2}$} \, ,\\
\, L + \frac{\ell}{2} - t \quad&\text{for $\quad \frac{L+\ell}{2} < t < \frac{2L+\ell}{2}$} \, , \\
\, 0 \quad&\text{for $\quad t >\frac{2L+\ell}{2}$} \, .
\end{cases} 
\end{align}
This result is true irrespective of whether $L > \ell$ or $L<\ell$.

Time evolution of mutual information of two disconnected region has also been studied for holographic CFTs. Unlike the mutual information in current dominated theories, the mutual information for holographic theories in the scaling limit ($\beta \ll t,L,\ell$) depends on whether $L>\ell$ or $L<\ell$. For $L<\ell$, the mutual information vanishes for all time, whereas for $L>\ell$, the mutual information in the thermal double model is given by \cite{gq-6}
\begin{align}
    I(A|B)(t) \, = \, 4 s_{eq} \, \times \label{eq-mi-holo}
\begin{cases}
\, 0 \quad&\text{for $\quad t <\frac{\ell}{2}$} \, ,\\[0ex]
\, t - \frac{\ell}{2} \quad&\text{for $\quad \frac{\ell}{2} < t < \frac{L}{2}$} \, ,\\
\, L - \frac{\ell}{2} - t \quad&\text{for $\quad \frac{L}{2} < t < \frac{2L-\ell}{2}$} \, , \\
\, 0 \quad&\text{for $\quad t >\frac{2L-\ell}{2}$} \, .
\end{cases} 
\end{align}

Another quantity that captures the entanglement between two disconnected regions is the logarithmic negativity, which is defined as \cite{Vidal:2002zz}
\begin{align}
    \E(A|B) \, \equiv \, \log \, \text{tr} \, |\rho_{AB}^{T_{B}}| \, ,
\end{align}
where $\rho_{A\cup B}^{T_{B}}$ denotes the partial transpose with respect to region $B$. The time evolution of negativity after a quench was studied in \cite{Coser:2014gsa} for theories for which we expect quasi-particle picture to be valid. It was found that the logarithmic negativity at any instant of time is proportional to the mutual information. More precisely, the negativity is given by
\begin{align}
    \E(A|B)(t) \, = \, \frac{3}{4} \, I(A|B)(t) \, .
\end{align}

Recently, a new measure of entanglement between two disconnected regions, called the reflected entropy, was introduced in \cite{faulkner}. This involves finding the `canonical' purification of a mixed state $\rho$. Consider a mixed state $\rho \in \h$ in its eigenbasis,
\begin{equation}
    \rho \, = \, \sum_{a} \rho_{a} \ket{\rho_{a}}\bra{\rho_{a}} \, .
\end{equation}
The canonical purification of this state is denoted by $\ket{\sqrt{\rho}} \in \h\otimes\h'$ and is given by
\begin{align}
    \ket{\sqrt{\rho}} \, = \, \sum_{a} \sqrt{\rho_{a}} \ket{\rho_{a}}\otimes\ket{\rho_{a}} \, .
\end{align}
For example, the canonical purification of a thermal state is the thermofield double state. 

Now given a density matrix $\rho_{AB} \, \in \, \h_{A}\otimes\h_{B} $ and its canonical purification $\ket{\sqrt{\rho_{AB}}} \, \in \, \h_{A}\otimes\h'_{A}\otimes\h_{B}\otimes\h'_{B} $, the reflected entropy is defined as
\begin{align}
    S_{R}(A|B) \, \equiv \, - \, \text{tr} \, \rho_{AA'} \, \log \rho_{AA'} \, \quad\quad \text{where} \quad\quad \, \rho_{AA'} \, = \, \text{tr}_{BB'} \ket{\sqrt{\rho_{AB}}}\bra{\sqrt{\rho_{AB}}} \, .
\end{align}

Our goal in this paper is to study the time evolution of the reflected entropy in rational and holographic CFTs\footnote{Time evolution of the reflected entropy after a local quench was studied in \cite{sr-local-1,sr-local-2}.}. Owing to its simplicity, we use the thermal double model to study this time evolution. 
For rational\footnote{Though, as we will discuss in Sec.~(\ref{sec-case-2}), our results for rational CFTs are valid for any current dominated CFT.} CFTs, we find that the time dependence of the reflected entropy of two disconnected regions (with arbitrary choice of $L$ and $\ell$) in the scaling limit is the same as the time dependence of the mutual information. That is,
\begin{align}
    S_{R}(A|B)(t) \, = \, I(A|B)(t) \, .\label{eq-res}
\end{align}
For holographic theories, we only focus in the limit $L\to\infty$ and finite $\ell$. In this case, we find that the time evolution of the reflected entropy is 
\begin{align}
    S_{R}(A|B) \, = \, 4 s_{eq} \, \times \label{eq-sr-holo-fin-intro}
\begin{cases}
\, 0 \quad&\text{for $\quad t <\frac{\ell}{2}$} \, ,\\[0ex]
\, t - \frac{\ell}{2} \quad&\text{for $\quad t > \frac{\ell}{2}$} \, . 
\end{cases}  
\end{align}

Various properties of the reflected entropy were derived in \cite{faulkner}. One such property is that the reflected entropy can never be less than the mutual information. That is,
\begin{align}
    S_{R}(A|B) \, \ge \, I(A|B) \, . \label{eq-sr-bound}
\end{align}
However, if a pure state  $\ket{\psi_{ABC}} \in \h_{A}\otimes\h_{B}\otimes\h_{C}$ has only bipartite entanglement, then it was recently shown in \cite{Akers:2019gcv} that the bound in Eq.~\eqref{eq-sr-bound} is saturated. If we take the quasi-particle picture for the evolution of entanglement seriously, then it suggests that the time-dependent state, in the scaling limit, has bipartite entanglement structure. Our result in Eq.~\eqref{eq-res} for rational theories provides some more evidence for this bipartite entanglement structure. (Note that our result does not prove the bipartite entanglement structure as GHZ states are also known to saturate the bound in Eq.~\eqref{eq-sr-bound}.) 

The rest of this paper is organized as follows. In Sec.~(\ref{sec-sr-cft}), we review the tools that we will use in this paper to calculate the time dependence of the reflected entropy. In particular, we review the replica trick for computing reflected entropy in Sec.~(\ref{sec-rep-trick}) and the holographic dual of the reflected entropy in Sec.~(\ref{sec-sr-ent-wedge}). We perform the main calculations in Sec.~(\ref{sec-quench-cft}) and in Sec.~(\ref{sec-holo-tdm}). In Sec.~(\ref{sec-quench-cft}), we mostly focus on rational CFTs and use the replica trick to calculate the time dependence of reflected entropy in a thermal double model. In Sec.~(\ref{sec-holo-tdm}), we focus on holographic CFTs and use the holographic formula for reflected entropy to calculate the time dependence of reflected entropy. We end with a summary and some possible extensions of our work in Sec.~(\ref{sec-disc}).

\section{Reflected entropy in CFTs} \label{sec-sr-cft} 

In this section, we briefly review the reflected entropy. In Sec.~(\ref{sec-rep-trick}), we review the replica trick approach of computing the reflected entropy. Then in Sec.~(\ref{sec-sr-ent-wedge}), we discuss the holographic dual of the reflected entropy. 

\subsection{Replica trick in $(1+1)$-dimensions} \label{sec-rep-trick}

A replica trick for computing reflected entropy was developed in \cite{faulkner}. This involves writing the reflected entropy in terms of correlation functions of certain codimension-$2$ twist operators inserted at the boundaries of regions $A$ and $B$. This trick is powerful especially in $(1+1)$-dimensions where the twist operators become local operators inserted at the end points of regions $A$ and $B$. In this section, we merely summarize this method of computing reflected entropy in $(1+1)$-dimensional CFTs and refer the readers to \cite{faulkner} for more details.

Reflected entropy, using replica trick, is given by\footnote{The order of limits may not commute. The correct order, as argued in \cite{sr-local-2}, is to first take $n\to 1$ and then take $m\to 1$.} 
\begin{align}
    S_{R}(A|B) \, = \, \lim_{m\to 1} \, \lim_{n\to 1} \, \frac{1}{1-n} \, \log \left(\frac{Z_{n,m}}{(Z_{1,m})^{n}}\right) \, , \label{eq-rep-sr}
\end{align}
where $Z_{n,m}$ is a correlation function of twist operators on CFT$^{\otimes mn}$. In particular, if we denote the end points of region $A$ by $a_{1}$ and $a_{2}$ and those of region $B$ by $b_{1}$ and $b_{2}$, then 
\begin{align}
    Z_{n,m} \, = \, \big\langle \, \sigma_{A}(a_{1}) \, \bar{\sigma}_{A}(a_{2}) \, \sigma_{B}(b_{1}) \, \bar{\sigma}_{B}(b_{2}) \, \big\rangle_{CFT^{\otimes mn}} \, . \label{eq-znm}
\end{align}
The conformal dimensions of these twist operators are \cite{faulkner}
\begin{align}
    h_{A} \, = \, h_{B} \, = \, n \, h_{m} \, ,
\end{align}
where 
\begin{align}
    h_{m} \, = \, \frac{c}{24} \, \frac{m^{2}-1}{m} \, , \label{eq-twist}
\end{align}
is the conformal dimension of the usual twist operators used in the calculation of the entanglement entropy \cite{Calabrese:2009qy}. 

We will use Eq.~\eqref{eq-rep-sr} in Sec.~(\ref{sec-quench-cft}) to study the time evolution of the reflected entropy in rational CFTs. We will find that the time-dependent reflected entropy, in the scaling limit, is governed by various operator product expansion (OPE) limits of twist operators in Eq.~\eqref{eq-znm}. Therefore, we now review the OPEs of twist operators in Eq.~\eqref{eq-znm}. The OPE of these operators, as discussed in \cite{faulkner}, is given by following fusion rules:
\begin{align}
    \sigma_{A} \, \bar{\sigma}_{A} \, \to \, \mathbf{1} \, \quad\quad\quad\quad \sigma_{B} \, \bar{\sigma}_{B} \, \to \, \mathbf{1} \, \quad\quad\quad\quad \sigma_{A} \, \bar{\sigma}_{B} \, \to \, \sigma_{AB} \, \label{eq-ope}
\end{align}
The conformal dimension of $\sigma_{AB}$ is given by
\begin{align}
    h_{AB} \, = \, 2 \, h_{n} \, ,
\end{align}
where $h_{n}$ is defined as in Eq.~\eqref{eq-twist}. Moreover, the OPE coefficient for the last fusion rule in Eq.~\eqref{eq-ope} is
\begin{align}
    C_{n,m} \, = \, (2m)^{-4h_{n}} \, .
\end{align}

This finishes our brief review of the replica trick method of computing the reflected entropy. Before we apply this method in Sec.~(\ref{sec-quench-cft}), we discuss the bulk dual of the reflected entropy for holographic CFTs.

\subsection{Holographic dual of reflected entropy} \label{sec-sr-ent-wedge}

In AdS-CFT correspondence, the bulk dual of a boundary subregion is the entanglement wedge. The entanglement wedge corresponding to a boundary subregion  is the bulk domain of dependence of a spacelike slice between that boundary subregion and its corresponding Hubeny-Rangamani-Takayanagi (HRT) surface. When the boundary subregion is the union of two disconnected subregion, the entanglement wedge can either be `connected' or `disconnected'. The connectedness of the entanglement wedge can be quantified using a bulk quantity, called the `entanglement wedge cross-section', which was defined in \cite{eop-1,eop-2}. In the following, we review this bulk quantity and its relation to the reflected entropy. 

The entanglement wedge cross-section for boundary regions $A$ and $B$, $E_{W}(A|B)$, can be defined as follows \cite{eop-1,eop-2}: Let us denote the HRT surfaces corresponding to boundary regions $A\cup B$ by $m_{AB}$ and the restriction of the entanglement wedge on some time slice by $M_{AB}$. Then the region $M_{AB}$ is such that 
\begin{align}
    \partial M_{AB} \, = \, A \, \cup \, B \, \cup \, m_{AB} \, .
\end{align}
Now let us divide $m_{AB}$ into two parts as
\begin{align}
    m_{AB} \, = \, m_{AB}^{(A)} \, \cup \, m_{AB}^{(B)} \, .\label{eq-div}
\end{align}
With this division, we define the entanglement wedge cross-section, $E_{W}(A|B)$, as
\begin{align}
    E_{W}(A|B) \, = \, \text{min}_{m_{A}} \, \frac{ \, \text{Area}\Big( \Sigma_{AB} \Big) \,}{4G} \, ,\label{eq-ent-wc}
\end{align}
where the minimization is over all possible divisions in Eq.~\eqref{eq-div} and where $\Sigma_{AB} \subset M_{AB}$ is such that
\begin{align}
    \partial \Sigma_{AB} \, = \, \partial \left(A \cup m_{AB}^{(A)}\right) \, = \, \partial \left(B \cup m_{AB}^{(B)}\right) \, ,
\end{align}
and it is homologous to $A\cup m_{AB}^{(A)}$ and $B\cup m_{AB}^{(B)}$.

By construction, $E_{W}(A|B)$ trivially vanishes when the entanglement wedge of $A \cup B$ is disconnected. In this sense, it is a measure of how connected the entanglement wedge is. 

It was argued and derived using the holographic replica trick in \cite{faulkner} that the boundary dual of the entanglement wedge cross-section is the reflected entropy. The precise relation between these quantities is \cite{faulkner}
\begin{align}
    S_{R}(A|B) \, = \, 2 \, E_{W}(A|B) \, .\label{eq-ref-ent-holo}
\end{align}
This relation is valid in any dimension and for any holographic state. This provides a useful tool to compute the reflected entropy for holographic states. We will use this formula in Sec.~(\ref{sec-holo-tdm}) to study the time-dependent reflected entropy in a holographic thermal double model.

\section{Time dependence of reflected entropy in rational CFTs} \label{sec-quench-cft}

Consider a doubled copy of a $(1+1)$-d CFT in a thermofield double state given in Eq.~\eqref{eq-tfd}. Let $A_{1}$ and $B_{1}$ are two disconnected regions in CFT$_{1}$ whereas $A_{2}$ and $B_{2}$ are their identical counterparts in CFT$_{2}$. We take regions $A$ and $B$ to be the union $A_{1}\cup A_{2}$ and $B_{1}\cup B_{2}$ respectively. The reduced density matrix of regions $A$ and $B$ can be constructed as a Euclidean path-integral over an infinite cylinder of size $\beta$ with open cuts above and below regions $A$ and $B$ \cite{hartman-maldacena}.

Now according to Eq.~\eqref{eq-rep-sr} and Eq.~\eqref{eq-znm}, the reflected entropy in the thermal double model is given in terms of the correlation function of twist operators in a cylinder. We follow \cite{hartman-maldacena,hartman-2} and insert the operators at the end points of regions $A$ and $B$ and at arbitrary Euclidean time. Then we analytically continue to Lorentzian time to get the time-dependence of the reflected entropy. 

If we take regions $A_{1}$ and $A_{2}$ to be intervals $[x_{1},x_{2}]$ and regions $B_{1}$ and $B_{2}$ to be intervals $[x_{3},x_{4}]$, then the reflected entropy at a given time is given in terms of the following correlation function:
\begin{align}
    Z_{n,m}^{\text{cyl}} \, = \,  \big\langle \, \sigma_{A}(z_{1},\bar{z}_{1})  \bar{\sigma}_{A}(z_{2},\bar{z}_{2})  \sigma_{B}(z_{3},\bar{z}_{3})  \bar{\sigma}_{B}(z_{4},\bar{z}_{4}) \sigma_{B}(z_{5},\bar{z}_{5})  \bar{\sigma}_{B}(z_{6},\bar{z}_{6})  \sigma_{A}(z_{7},\bar{z}_{7})  \bar{\sigma}_{A}(z_{8},\bar{z}_{8}) \, \big\rangle^{\text{cyl}}_{CFT^{\otimes mn}} \, . \label{eq-znm-cyl}
\end{align}
In this correlation function,
\begin{align}
    z_{i} \, = \, x_{i} - t - i {\beta}/{4} \, , \quad\quad\quad\quad \bar{z}_{i} \, = \, x_{i} + t + i {\beta}/{4} \, ,\label{eq-zi}
\end{align}
for $i \, = \, \{1,2,3,4\}$, and 
\begin{align}
    z_{i} \, = \, \bar{z}_{9-i} \, \quad\quad\quad\quad \bar{z}_{i} \, = \, z_{9-i} \, ,\label{eq-zi-2} 
\end{align}
for $i \, = \, \{5,6,7,8\}$. Note that $z^{*}_{i} \ne \bar{z}_{i}$ is due to the analytic continuation to Lorentzian time as discussed above. 

Note that an infinite cylinder can be mapped to a complex plane using the following conformal transformation:
\begin{align}
    w = \exp\left(2\pi z/\beta\right) \, \quad\quad\quad \bar{w}= \exp\left(2\pi \bar{z}/\beta\right) \, .
\end{align}
Using this conformal transformation, we write the correlation function in Eq.~\eqref{eq-znm-cyl} as a correlation function on a plane. This yields
\begin{align}
    Z_{n,m}^{\text{cyl}} \, = \,  \left(\frac{2\pi}{\beta}\right)^{16 n h_{m}} \, \big|w_{1}w_{2}w_{3}w_{4}w_{5}w_{6}w_{7}w_{8}\big|^{2nh_{m}} \, Z_{n,m}^{\text{plane}} \, , \label{eq-znm-cyl-2}
\end{align}
where $Z_{n,m}^{\text{plane}}$ is the following correlation function on a plane:
\begin{align}
        \big\langle \, \sigma_{A}(w_{1},\bar{w}_{1})  \bar{\sigma}_{A}(w_{2},\bar{w}_{2})  \sigma_{B}(w_{3},\bar{w}_{3})  \bar{\sigma}_{B}(w_{4},\bar{w}_{4})  \sigma_{B}(w_{5},\bar{w}_{5})  \bar{\sigma}_{B}(w_{6},\bar{w}_{6})  \sigma_{A}(w_{7},\bar{w}_{7})  \bar{\sigma}_{A}(w_{8},\bar{w}_{8}) \, \big\rangle^{\text{plane}}_{CFT^{\otimes mn}} \, . \label{eq-znm-plane}
\end{align}

Now recall from Eq.~\eqref{eq-rep-sr} that the reflected entropy is given in terms of the ratio $Z_{n,m}^{\text{cyl}}/\big(Z_{1,m}^{\text{cyl}}\big)^{n}$. We find that the conformal factor in Eq.~\eqref{eq-znm-cyl-2} drops out from this ratio, and we get
\begin{align}
    \frac{Z_{n,m}^{\text{cyl}}}{\big(Z_{1,m}^{\text{cyl}}\big)^{n}} \, = \, \frac{Z_{n,m}^{\text{plane}}}{\big(Z_{1,m}^{\text{plane}}\big)^{n}} \, .
\end{align}
This is an interesting observation as it implies that the conformal factor in Eq.~\eqref{eq-znm-cyl-2} does not contribute to the reflected entropy\footnote{In fact, the conformal factor drops out from Eq.~\eqref{eq-rep-sr} even before taking the replica limit. This means that the conformal factor does not contribute to the Renyi generalization of the reflected entropy as well.}. Moreover, the reflected entropy in the thermal double model is given by
\begin{align}
    S_{R}(A|B)(t) \, = \, \lim_{m\to 1} \, \lim_{n\to 1} \, \frac{1}{1-n} \, \log \left(\frac{Z^{\text{plane}}_{n,m}}{(Z^{\text{plane}}_{1,m})^{n}}\right) \, . \label{eq-sr-tdm}
\end{align}
In the following, we compute the the time dependence of $Z_{n,m}^{\text{plane}}$ and then combine it with Eq.~\eqref{eq-sr-tdm} to get the time dependence of the reflected entropy.

\subsection{Setup} \label{sec-tdm-calc}

The discussion in the previous subsection was for arbitrary regions $A$ and $B$. From now on, for concreteness, we take regions $A_{1}$, $B_{1}$, $A_{2}$, and $B_{2}$ to be of equal size $L$. Furthermore, we denote the separation between $A_{1}$ ($A_{2}$) and $B_{1}$ ($B_{2}$) by $\ell$. More precisely, we choose $x_{1}$, $x_{2}$, $x_{3}$, and $x_{4}$ in Eq.~\eqref{eq-zi} to be
\begin{align}
    x_{1} \, = \, -L -\ell/2 \, , \quad\quad x_{2} \, = \, -\ell/2 \, , \quad\quad x_{3} \, = \, \ell/2 \, , \quad\quad x_{4} \, = \, L+\ell/2 \, .\label{eq-xi}
\end{align}

Having specified regions $A$ and $B$, we now compute the time-dependent reflected entropy. In the following, we consider the following three cases separately:
\begin{itemize}
    \item Case $1$: $L\to \infty$. 
    \item Case $2$: $L > \ell$.
    \item Case $3$: $L < \ell$. 
\end{itemize}

\subsection{Case $1$: $L \to \infty$} \label{sec-case-1}

This case is a simplified version of case $2$. However, we still think it is a good idea to consider it separately. This is because we expect this simpler case to shed light on interesting aspects of the calculation that will help us in studying case $2$ and case $3$. More importantly, as we will see in this section, the time dependence of the reflected entropy in this case is completely fixed by the conformal symmetry. Therefore, the results of this section are valid for \textit{all} CFTs. 

In this case, $Z_{n,m}^{\text{plane}}$ is given by a four-point function on the plane
\begin{align}
    Z^{\text{plane}}_{n,m} \, = \, \big\langle \, \sigma_{A}(w_{1},\bar{w}_{1})  \bar{\sigma}_{B}(w_{2},\bar{w}_{2})  \sigma_{B}(w_{3},\bar{w}_{3})  \bar{\sigma}_{A}(w_{4},\bar{w}_{4}) \, \big\rangle^{\text{plane}}_{CFT^{\otimes mn}} \, , \label{eq-znm-case1}
\end{align}
where 
\begin{align}
    w_{1} \, = \, -i e^{-\frac{2\pi}{\beta}(t+\ell/2)} \, , \quad\quad\quad\quad \bar{w}_{1} \, = \, i e^{\frac{2\pi}{\beta}(t-\ell/2)} \, , \label{eq-w1-c1}\\
    w_{2} \, = \, -i e^{-\frac{2\pi}{\beta}(t-\ell/2)} \, , \quad\quad\quad\quad \bar{w}_{2} \, = \, i e^{\frac{2\pi}{\beta}(t+\ell/2)} \, , \label{eq-w2-c1}
\end{align}
and $w_{3} \, = \, \bar{w}_{2} \, $, $\bar{w}_{3} \, = \, {w}_{2} \, $, ${w}_{4} \, = \, \bar{w}_{1} \, $, and  $\bar{w}_{4} \, = \, {w}_{1} \, $.

Recall that conformal symmetry fixes the four-point function on a plane up to a unknown function of the cross-ratio. Let us consider the following cross-ratio:
\begin{align}
    \eta \, = \, \bar{\eta} \, \equiv \, \frac{(w_{1}-\w_{1}) (w_{2}-\w_{2}) }{(w_{1}-\w_{2}) (w_{2}-\w_{1}) } \, .\label{eq-eta}
\end{align}
Now using Eqs.~\eqref{eq-w1-c1}-\eqref{eq-w2-c1}, we get
\begin{align}
    \eta \, = \, \frac{2 \, \sinh^{2}(2\pi t/\beta)}{\cosh(4\pi t/\beta) \, + \, \cosh(2\pi \ell/\beta)} \, .
\end{align}
In the scaling limit, that is $\beta \to 0$ limit, this expression simplifies to
\begin{align}
    \eta \, = \, \frac{1}{1 \, + \, \exp\left(-\frac{2\pi}{\beta} (2t-\ell)\right)} \, = \, \begin{cases}
\, 0 \quad&\text{for $\quad t <\frac{\ell}{2}$} \, ,\\[0ex]
\, 1 \quad&\text{for $\quad t>\frac{\ell}{2}$} \, ,
\end{cases} \, . \label{eq-eta-scaling}
\end{align}
Note that $\eta \to 0$ corresponds to the OPE limit
\begin{align}
    (w_{1},\w_{1}) \leftrightarrow (w_{4},\w_{4}) \quad\quad \text{and} \quad\quad (w_{2},\w_{2}) \leftrightarrow (w_{3},\w_{3}) \, ,\label{eq-ope-c1-1}
\end{align}
whereas
$\eta \to 1$ corresponds to the OPE limit
\begin{align}
    (w_{1},\w_{1}) \leftrightarrow (w_{2},\w_{2}) \quad\quad \text{and} \quad\quad (w_{3},\w_{3}) \leftrightarrow (w_{4},\w_{4}) \, .\label{eq-ope-c1-2}
\end{align}
This means that the time dependence of the reflected entropy is governed by one of the OPEs in Eqs.~\eqref{eq-ope}. 

We now use the above observation to compute $Z^{\text{plane}}_{n,m}$ in Eq.~\eqref{eq-znm-case1} as a function of time. For $t<\ell/2$, we take the OPE limit in Eq.~\eqref{eq-ope-c1-1} to get
\begin{align}
    Z^{\text{plane}}_{n,m} \, = \, |w_{1}-w_{4}|^{-4nh_{m}} \, |w_{2}-w_{3}|^{-4nh_{m}} \, .
\end{align}
This implies, 
\begin{align}
     \frac{Z^{\text{plane}}_{n,m}}{(Z^{\text{plane}}_{1,m})^{n}} \, = \, 1 \, ,
\end{align}
and hence, by virtue of Eq.~\eqref{eq-sr-tdm},
\begin{align}
    S_{R}(A|B) \, = \, 0 \, .
\end{align}
For $t>\ell/2$, on the other hand, we take the OPE limit in Eq.~\eqref{eq-ope-c1-2} to get
\begin{align}
    Z^{\text{plane}}_{n,m} \, =& \, (2m)^{-8 \, h_{n}} \,  |w_{1}-w_{2}|^{-4nh_{m} + 4h_{n}} \, |w_{3}-w_{4}|^{-4nh_{m}+4h_{n}} \, |w_{1}-w_{4}|^{-4h_{n}} \, |w_{2}-w_{3}|^{-4h_{n}} \,  ,\\
    =& \, (2m)^{-8 \, h_{n}} \,  |w_{1}-w_{2}|^{-4nh_{m}} \, |w_{3}-w_{4}|^{-4nh_{m}} \, \left( \frac{1-\eta}{\eta} \right)^{4h_{n}} \,  ,
\end{align}
where we have used Eq.~\eqref{eq-eta}. This implies
\begin{align}
     \log \, \frac{Z^{\text{plane}}_{n,m}}{(Z^{\text{plane}}_{1,m})^{n}} \, = \, -8 h_{n} \log (2m) + 4h_{n} \, \log \left( \frac{1-\eta}{\eta} \right)  \, .
\end{align}
In the scaling limit, this becomes
\begin{align}
     \log \, \frac{Z^{\text{plane}}_{n,m}}{(Z^{\text{plane}}_{1,m})^{n}} \, = \, - \, \frac{16\pi h_{n}}{\beta}  \, \left( t - \ell/2 \right)  \, .
\end{align}
Combinging this result with Eq.~\eqref{eq-sr-tdm}, we get
\begin{align}
    S_{R}(A|B) \, = \, 4 \, s_{eq} \, \left( t - \ell/2\right) \, ,
\end{align}
where $s_{eq}$ is given in Eq.~\eqref{eq-seq}.

To summarize, the time-dependence of the reflected entropy in the $L\to\infty$ limit is given by 
\begin{align}
    S_{R}(A|B)(t) = 4  s_{eq} \times \label{eq-fin-case-1-gen}
\begin{cases}
\, 0 \quad&\text{for $\quad t <\frac{\ell}{2}$} \, ,\\[0ex]
\, t - \frac{\ell}{2} \quad&\text{for $\quad t>\frac{\ell}{2}$} \, .
\end{cases}
\end{align}
This is the main result of this section. We reiterate that that this result was fixed by the OPEs of the twist operators. We did not have to assume anything about the spectrum of the CFT. Hence, this result is valid for \textit{all} CFTs, as advertised at the beginning of this section. 

\subsection{Case $2$: $L > \ell$} \label{sec-case-2}

Recall from Eq.~\eqref{eq-znm-plane} that $Z_{n,m}^{\text{plane}}$ is a eight-point function on a plane. 
The operators in this correlation function are inserted at
\begin{align}
    w_{i} \, = \w_{9-i} \, = \, -i e^{-\frac{2\pi}{\beta}(t-x_{i})} \, , \quad\quad\quad\quad \bar{w}_{i} \, = \, w_{9-i} \, i e^{\frac{2\pi}{\beta}(t+x_{i})} \, , 
\end{align}
for $i \, = \, \{1,2,3,4\}$ and $x_{i}$ are given in Eq.~\eqref{eq-xi}. 

The calculation of an eight-point function, in general, is quite difficult. However, as we saw in the last subsection, the operators approach each other in the scaling limit. To decide which two operators approach each other, we follow \cite{hartman-2} and consider the following cross-ratios:
\begin{align}
    \eta_{ij} \, = \, \bar{\eta}_{ij} \, \equiv \, \frac{(w_{i}-\w_{i}) (w_{j}-\w_{j}) }{(w_{i}-\w_{j}) (w_{j}-\w_{i}) } \, .\label{eq-eta-ij}
\end{align}
In the scaling limit, these cross-ratios become
\begin{align}
    \eta_{ij} \, = \, \frac{1}{1 \, + \, \exp\left(-\frac{2\pi}{\beta} (2t-|x_{i}-x_{j}|)\right)} \, = \, \begin{cases}
\, 0 \quad&\text{for $\quad t <\frac{|x_{i}-x_{j}|}{2}$} \, ,\\[0ex]
\, 1 \quad&\text{for $\quad t>\frac{|x_{i}-x_{j}|}{2}$} \, ,
\end{cases} \, . \label{eq-eta-ij-scaling}
\end{align}
These cross-ratios are sufficient to determine the correct OPE limit in the Euclidean signature. However, this is not the case in the Lorentzian signature, as was discussed in \cite{hartman-2}. This is because an operator can approach a light cone of some other operator. To remedy this, we follow \cite{hartman-2} and consider two more cross-ratios
\begin{align}
\xi \, \equiv& \, \frac{(w_{1}-w_{2}) (w_{5}-w_{6}) }{(w_{1}-w_{5}) (w_{2}-w_{6}) } \, = \, \frac{(\w_{8}-\w_{7}) (\w_{4}-\w_{3}) }{(\w_{8}-\w_{4}) (\w_{7}-\w_{3}) } \, ,\\
\bar{\xi} \equiv& \, \frac{(w_{8}-w_{7}) (w_{4}-w_{3}) }{(w_{8}-w_{4}) (w_{7}-w_{3}) } \, = \, \frac{(\w_{1}-\w_{2}) (\w_{5}-\w_{6}) }{(\w_{1}-\w_{5}) (\w_{2}-\w_{6}) } \, .
\end{align}
In the scaling limit, these cross-ratios become
\begin{align}
\xi \, = \,  \exp\left(-\frac{2\pi}{\beta} (2t+\ell)\right) \, \to \, 0 \, , \label{eq-xi-sc}
\end{align}
and   
\begin{align}
\bar{\xi} \, = \, \frac{1}{1 \, + \, \exp\left(\frac{2\pi}{\beta} (|2t-L-\ell| - L)\right)} \, = \, \begin{cases}
\, 0 \quad&\text{for $\quad t <\frac{\ell}{2}$} \, ,\\[0ex]
\, 1 \quad&\text{for $\quad \frac{\ell}{2} < t < \frac{2L+\ell}{2}$} \, ,\\
\, 0 \quad&\text{for $\quad t >\frac{2L+\ell}{2}$} \, ,\\[0ex] 
\end{cases} \, . \label{eq-xi-bar-sc}
\end{align}

Now we calculate the time-dependence of $Z_{n,m}^{\text{plane}}$ and that of $S_{R}(A|B)$. 
For $t < \ell/2$, all the cross-ratios vanishes. This suggests that the following points approach each other:
\begin{align}
w_{1} \leftrightarrow& \, w_{8} \, \quad\quad w_{2} \leftrightarrow w_{7} \, \quad\quad \, w_{3} \leftrightarrow w_{6} \, \quad\quad w_{4} \leftrightarrow w_{5} \, ,\label{eq-conf-1-1}\\
\w_{1} \leftrightarrow& \, \w_{8} \, \quad\quad \w_{2} \leftrightarrow \w_{7} \, \quad\quad \, \w_{3} \leftrightarrow \w_{6} \, \quad\quad \w_{4} \leftrightarrow \w_{5} \, . \label{eq-conf-1-2}
\end{align}
In this limit, we can use OPEs in Eq.~\eqref{eq-ope} to write $Z_{n,m}^{\text{plane}}$ as
\begin{align}
Z_{n,m}^{\text{plane}} \, = \, |w_{1}-w_{8}|^{-2nh_{m}} \, |w_{2}-w_{7}|^{-2nh_{m}} \, |w_{3}-w_{6}|^{-2nh_{m}} \, |w_{4}-w_{5}|^{-2nh_{m}}  \, .
\end{align} 
Now we insert this in Eq.~\eqref{eq-sr-tdm} and find
\begin{align}
S_{R}(A|B) \, = \, 0 \, . \label{eq-sr-c2-1}
\end{align}

Now let us consider $\ell/2 < t < L/2$. In this case, $\eta_{23}$ and $\bar{\xi}$ approach to $1$ while all other cross-ratios vanishes. This corresponds to the  following configuration 
\begin{align}
w_{1} \leftrightarrow& \, w_{8} \, \quad\quad w_{2} \leftrightarrow w_{3} \, \quad\quad \, w_{4} \leftrightarrow w_{5} \, \quad\quad w_{6} \leftrightarrow w_{7} \, ,\label{eq-conf-2-1}\\
\w_{1} \leftrightarrow& \, \w_{8} \, \quad\quad \w_{2} \leftrightarrow \w_{3} \, \quad\quad \, \w_{4} \leftrightarrow \w_{5} \, \quad\quad \w_{6} \leftrightarrow \w_{7} \, .\label{eq-conf-2-2}
\end{align}
In this limit, $Z_{n,m}^{\text{plane}}$ is again fixed by OPEs in Eq.~\eqref{eq-ope}. By using these OPE relations, we get
\begin{align}
Z_{n,m}^{\text{plane}} \, = \, &(2m)^{-8h_{n}} \,  |w_{1}-w_{8}|^{-2nh_{m}} \, |w_{2}-w_{3}|^{-2nh_{m}} \, |w_{4}-w_{5}|^{-2nh_{m}} \, |w_{6}-w_{7}|^{-2nh_{m}}  \nonumber\\ &\, \times \left(\frac{|w_{2}-w_{3}|}{|w_{2}-w_{7}||\w_{3}-\w_{6}|} \right)^{4h_{n}} \, .
\end{align} 
This implies
\begin{align}
    \frac{Z^{\text{plane}}_{n,m}}{(Z^{\text{plane}}_{1,m})^{n}} \, = \, (2m)^{-8h_{n}} \, \left( \frac{1 - \eta_{23}}{\eta_{23}} \right)^{4h_{n}} \, .
\end{align}
Now using Eq.~\eqref{eq-sr-tdm} and taking the scaling limit, we get
\begin{align}
S_{R}(A|B) \, = \, 4 \, s_{eq} \, \left( t - \ell/2\right)  \, .
\end{align}

Now we consider $L/2 < t < (L+\ell)/2$. In this case, cross-ratios satisfy
\begin{align}
\eta_{12} \, , \, \eta_{34} \, , \, \eta_{23} \, , \, \bar{\xi} \, \to \, 1 \, . \label{eq-cr-dip-1}
\end{align}
and
\begin{align}
\eta_{13} \, , \, \eta_{24} \, , \, \eta_{14} \, , \, {\xi} \, \to \, 0 \, . \label{eq-cr-dip-2}
\end{align}
This implies the following configuration:
\begin{align}
w_{1} \leftrightarrow& \, w_{2} \, \quad\quad w_{3} \leftrightarrow w_{8} \, \quad\quad \, w_{4} \leftrightarrow w_{7} \, \quad\quad w_{5} \leftrightarrow w_{6} \, ,\label{eq-conf-3-1}\\
\w_{1} \leftrightarrow& \, \w_{6} \, \quad\quad \w_{2} \leftrightarrow \w_{5} \, \quad\quad \, \w_{3} \leftrightarrow \w_{4} \, \quad\quad \w_{7} \leftrightarrow \w_{8} \, .\label{eq-conf-3-2}
\end{align}
In this configuration, $w$'s and $\w$'s are in a different channel, and hence, this configuration does not correspond to any OPE limit. Therefore, the OPEs do not fix the eight point function in general. Nevertheless, for current dominated theories such as rational theories, we can treat the `left-movers' and `right-movers' separately \cite{hartman-2}. This allows us to choose different OPE channels for $w$'s and $\w$'s. Using this observation, we get
\begin{align}
Z_{n,m}^{\text{plane}} \, = \, (2m)^{-8h_{n}} \, &(w_{1}-w_{2})^{-2nh_{m}} \, (w_{3}-w_{8})^{-2nh_{m}+2h_{n}} \, (w_{4}-w_{7})^{-2nh_{m}+2h_{n}} \, (w_{5}-w_{6})^{-2nh_{m}} \nonumber\\
\times & (\w_{1}-\w_{6})^{-2nh_{m}+2h_{n}} \, (\w_{2}-\w_{5})^{-2nh_{m}+2h_{n}} \, (\w_{3}-\w_{4})^{-2nh_{m}} \, (\w_{7}-\w_{8})^{-2nh_{m}} \nonumber\\
\times& (w_{3}-w_{4})^{-2h_{n}} \, (w_{8}-w_{7})^{-2h_{n}} \, (\w_{1}-\w_{2})^{-2h_{n}} \, (\w_{6}-\w_{5})^{-2h_{n}} \, .
\end{align}
This implies 
\begin{align}
    \frac{Z^{\text{plane}}_{n,m}}{(Z^{\text{plane}}_{1,m})^{n}} \, = \, (2m)^{-8h_{n}} \, \left( \frac{1 - \bar{\xi}}{\bar{\xi}} \right)^{4h_{n}} \, . \label{eq-znm-grow}
\end{align}
Now using Eq.~\eqref{eq-sr-tdm} and using
\begin{align}
    \log \left(\frac{1-\bar{\xi}}{\bar{\xi}}\right) \, = \, \frac{2\pi}{\beta} \, \left( |2t-L-\ell| \, - \, L \right) \, ,
\end{align}
we get 
\begin{align}
S_{R}(A|B) \, = \, 4 \, s_{eq} \, \left( t - \ell/2\right)  \, .\label{eq-sr-dip-1}
\end{align}

Now we focus on $(L+\ell)/2 < t < (2L+\ell)/2$. In this case, $\eta_{14}$ and $\xi$ vanishes whereas all other cross-ratios approach $1$. This corresponds to the configuration
\begin{align}
w_{1} \leftrightarrow& \, w_{2} \, \quad\quad w_{3} \leftrightarrow w_{8} \, \quad\quad \, w_{4} \leftrightarrow w_{7} \, \quad\quad w_{5} \leftrightarrow w_{6} \, ,\label{eq-cr-grow-1}\\
\w_{1} \leftrightarrow& \, \w_{6} \, \quad\quad \w_{2} \leftrightarrow \w_{5} \, \quad\quad \, \w_{3} \leftrightarrow \w_{4} \, \quad\quad \w_{7} \leftrightarrow \w_{8} \, .\label{eq-cr-grow-2}
\end{align}
Note that this configuration is the same as that in Eqs.~\eqref{eq-conf-3-1}-\eqref{eq-conf-3-2}. This means that $Z^{\text{plane}}_{n,m}$ for rational theories still satisfies Eq.~\eqref{eq-znm-grow} and hence, the reflected entropy is given by
\begin{align}
S_{R}(A|B) \, = \, 4 \, s_{eq} \, \left( L + \ell/2 - t \right)  \, . \label{eq-sr-grow}
\end{align}

Finally, we consider $t> (2L+\ell)/2$. In this case, all $\eta_{ij} \to 1$ but $\xi = \bar{\xi} \, = \, 0  \, $. This corresponds to the configuration:
\begin{align}
w_{1} \leftrightarrow& w_{2} \, \quad\quad w_{3} \leftrightarrow w_{4} \, \quad\quad \, w_{5} \leftrightarrow w_{6} \, \quad\quad w_{7} \leftrightarrow w_{8} \, ,\label{eq-conf-4-1}\\
\w_{1} \leftrightarrow& \w_{2} \, \quad\quad \w_{3} \leftrightarrow \w_{4} \, \quad\quad \, \w_{5} \leftrightarrow \w_{6} \, \quad\quad \w_{7} \leftrightarrow \w_{8} \, .\label{eq-conf-4-2}
\end{align}
In this limit,  $Z_{n,m}^{\text{plane}}$ is once again fixed by the OPEs in Eq.~\eqref{eq-ope}. Using these OPEs, we get
\begin{align}
Z_{n,m}^{\text{plane}} \, = \, |w_{1}-w_{2}|^{-2nh_{m}} \, |w_{3}-w_{4}|^{-2nh_{m}} \, |w_{5}-w_{6}|^{-2nh_{m}} \, |w_{7}-w_{8}|^{-2nh_{m}} \, .
\end{align} 
Now using Eq.~\eqref{eq-sr-tdm}, we get
\begin{align}
S_{R}(A|B) \, = \, 0 \, . \label{eq-sr-late}
\end{align}

To summarize, we find that the reflected entropy is given by
\begin{align}
S_{R}(A|B) \, = \, 4 s_{eq} \, \times \label{eq-sr-2-fin}
\begin{cases}
\, 0 \quad&\text{for $\quad t <\frac{\ell}{2}$} \, ,\\[0ex]
\, t - \frac{\ell}{2} \quad&\text{for $\quad \frac{\ell}{2} < t < \frac{L+\ell}{2}$} \, ,\\
\, L + \frac{\ell}{2} - t \quad&\text{for $\quad \frac{L+\ell}{2} < t < \frac{2L+\ell}{2}$} \, , \\
\, 0 \quad&\text{for $\quad t >\frac{2L+\ell}{2}$} \, .
\end{cases}  \, .
\end{align}
Note that this precisely matches the time-evolution of the mutual information in Eq.~\eqref{eq-mi}. That is, 
\begin{align}
    S_{R}(A|B)(t) \, = \, I(A|B)(t) \,  \label{eq-sr-mi}
\end{align}
in a thermal double model. 

In the next subsection, we will see that this interesting result is valid even if $L < \ell$. 

\subsection{Case $3$: $L < \ell$}

Here we repeat the calculation of the eight-point function, $Z_{n,m}^{\text{plane}}$, and the reflected entropy but for $L < \ell$. Note that the time-dependence of cross-ratios that we derived in the last subsection, that is Eq.~\eqref{eq-eta-ij-scaling}, Eq.~\eqref{eq-xi-sc}, and Eq.~\eqref{eq-xi-bar-sc}, is still valid in this case. Therefore, we can use these results to decide which two points are approaching each other in the scaling limit. 

For $t< L/2$, all the cross-ratios vanishes. This corresponds to the same configurations as in Eqs.~\eqref{eq-conf-1-1}-\eqref{eq-conf-1-2}. This means that the reflected entropy is fixed by the OPEs and is the same as in Eq.~\eqref{eq-sr-c2-1}:
\begin{align}
S_{R}(A|B) \, = \, 0 \, .
\end{align}

Now we consider $L/2 < t <\ell/2$. In this case,  $\eta_{12}$ and $\eta_{34}$ approach to $1$ while all other cross-ratios vanishes. This corresponds to the  following configuration
\begin{align}
w_{1} \leftrightarrow& w_{2} \, \quad\quad w_{3} \leftrightarrow w_{4} \, \quad\quad \, w_{5} \leftrightarrow w_{6} \, \quad\quad w_{7} \leftrightarrow w_{8} \, ,\\
\w_{1} \leftrightarrow& \w_{2} \, \quad\quad \w_{3} \leftrightarrow \w_{4} \, \quad\quad \, \w_{5} \leftrightarrow \w_{6} \, \quad\quad \w_{7} \leftrightarrow \w_{8} \, .
\end{align}
Note that this configuration is the same as in Eqs.~\eqref{eq-conf-4-1}-\eqref{eq-conf-4-2}. This means that the reflected entropy vanishes as in Eq.~\eqref{eq-sr-late}:
\begin{align}
S_{R}(A|B) \, = \, 0 \, .
\end{align}

Now we focus on $\ell/2 < t < (L+\ell)/2$. In this case, cross-ratios have the same limits as in Eqs.~\eqref{eq-cr-dip-1}-\eqref{eq-cr-dip-2}, and hence, we have the same configuration as in Eqs.~\eqref{eq-conf-3-1}-\eqref{eq-conf-3-2}. Therefore, we deduce that the reflected entropy is the same as in Eq.~\eqref{eq-sr-dip-1}:
\begin{align}
S_{R}(A|B) \, = \, 4 \, s_{eq} \, \left( t - \ell/2\right) \, .
\end{align}

Now we assume that $(L+\ell)/2 < t < (2L+\ell)/2$. In this case, $\eta_{14}$ and $\xi$ vanishes whereas all other cross-ratios approach $1$. This corresponds to the same  configuration as in Eqs.~\eqref{eq-cr-grow-1}-\eqref{eq-cr-grow-2}. Therefore, we deduce that the reflected entropy is the same as in Eq.~\eqref{eq-sr-grow}:
\begin{align}
S_{R}(A|B) \, = \, 4 \, s_{eq} \, \left( L + \ell/2 - t \right) \, .
\end{align}

Finally, we consider $t>(2L+\ell)/2$. This again corresponds to the configurations in Eqs.~\eqref{eq-conf-4-1}-\eqref{eq-conf-4-2}. Therefore, we deduce that the reflected entropy vanishes:
\begin{align}
S_{R}(A|B) \, = \, 0 \, .
\end{align}

Combining the above results, we find that the time-dependent reflected entropy, even for $L < \ell$, is given by Eq.~\eqref{eq-sr-2-fin}. Hence, in a thermal double model, 
\begin{align}
S_{R}(A|B)(t) \, = \, I(A|B)(t) \, .\label{eq-sr-mi}
\end{align}

As we discussed in Sec.~(\ref{intro}), our result that the reflected entropy equals mutual information in a thermal double model provides some more evidence for the quasi-particle picture for the spread of entanglement. 

\section{Time dependence of reflected entropy in holographic CFTs} \label{sec-holo-tdm}

In this section, we compute the time evolution of the reflected entropy in the thermal double model using AdS-CFT correspondence. The holographic dual of the entangled state in Eq.~\eqref{eq-tfd} is a two-sided black brane \cite{Maldacena:2001kr}. This bulk geometry has two exterior regions, each corresponding to a single copy of the CFT. Therefore, the holographic dual of the thermal double model is a two-sided black brane where time is taken to run forwards on both of the exterior regions \cite{hartman-maldacena}. 

Since our focus in this work is only on $(1+1)$-dimensional CFTs, we consider the BTZ black brane in this section. The metric of the BTZ black string is
\begin{align}
    ds^{2} \, = \, - \frac{4\pi^{2}}{\beta^{2}} \, \sinh^{2}\rho \, dt^{2} \, + \, d\rho^{2} \, + \, \frac{4\pi^{2}}{\beta^{2}} \, \cosh^{2}\rho \, dx^{2} \, .
\end{align}
Note that the two exterior regions are related to each other by continuation $t \to t + i\beta/2$.

The BTZ black brane is locally equivalent to the Poincare AdS$_{3}$ spacetime \cite{Banados:1992wn}:
\begin{align}
    ds^{2} \, = \, \frac{1}{z^{2}} \, \left( dz^{2} - dx_{0}^{2}  + dx_{0}^{2}  \right) \, .\label{eq-met-poin}
\end{align}
Note that the two asymptotic boundaries of an eternal BTZ black brane corresponds to two Rindler wedges of the boundary of the Poincare AdS$_{3}$ \cite{Maldacena:1998bw,Parikh:2012kg}. Moreover, a point on an exterior region of the BTZ black brane can be mapped to a point on the Poincare AdS$_{3}$ using \cite{hartman-maldacena}
\begin{align}
    z \, =& \, \frac{1}{\cosh\rho} \, e^{2\pi x/\beta}  \, ,\label{eq-z-1}\\
    x_{1} \, =& \, \tanh\rho \, \cosh\left(2\pi t/\beta\right) \, e^{2\pi x/\beta}  \, ,\label{eq-x1-1}\\
    x_{0} \, =& \, \tanh\rho \, \sinh\left(2\pi t/\beta\right) \, e^{2\pi x/\beta}  \, \label{eq-x0-1}.
\end{align}
Since the two exterior regions of the BTZ black brane are related by the continuation $t \to t + i\beta/2$, we deduce that map between the other exterior region and the Poincare AdS$_{3}$ is 
\begin{align}
    z \, =& \, \frac{1}{\cosh\rho} \, e^{2\pi x/\beta}  \, ,\label{eq-z-2}\\
    x_{1} \, =& \, - \, \tanh\rho \, \cosh\left(2\pi t/\beta\right) \, e^{2\pi x/\beta}  \, ,\label{eq-x1-2}\\
    x_{0} \, =& \, - \, \tanh\rho \, \sinh\left(2\pi t/\beta\right) \, e^{2\pi x/\beta}  \, .\label{eq-x0-2}
\end{align}

In the following, we will use these maps to relate the calculation of the entanglement wedge cross-section to the length of a geodesic in the Poincare AdS$_{3}$ geometry. Before we discuss the entanglement wedge cross-section, we need to discuss how does the entanglement wedge of boundary regions $A$ and $B$ changes as a function of time. This is the topic of the next subsection.

\subsection{Time dependence of the entanglement wedge}

Here we consider the same setup as in Sec.~(\ref{sec-case-1}). That is, both regions $A$ and $B$ consist of identical semi-infinite intervals\footnote{An error was pointed out to us by Jonah Kudler-Flam and Yuya Kusuki in our analysis of finite size regions in an earlier version of this paper, and hence, the analysis has been removed from this version.} on each of the asymptotic boundaries of the BTZ black brane, and that they are separated by an interval of size $\ell$.

To understand what is the entanglement wedge for these boundary regions, we first need to understand the HRT surface for these boundary regions. The boundary anchored surfaces in the two-sided black brane were studied in \cite{hartman-maldacena}. It was found that the boundary anchored extremal surface can either go through the black hole from one asymptotic region to another or it can remain entirely in the exterior region. The area of the former surfaces grow linearly with time whereas the area of the latter surfaces scales as the size of the boundary regions where the surfaces are anchored. This implies that the correct HRT surface at any instant of time is given by one of the two possible configurations of the boundary anchored surfaces that we now discuss.

\begin{figure}
    \centering
    \includegraphics[scale=.65]{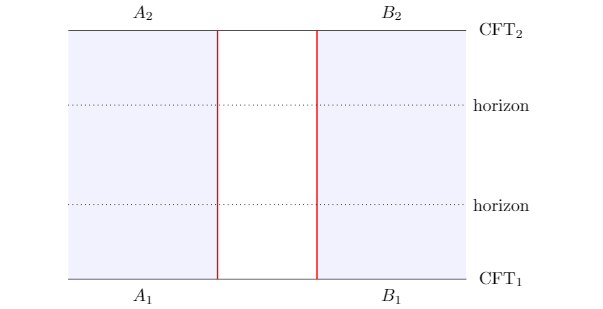}
    \caption{The pictorial representation of one of the two possibilities of the HRT surfaces corresponding to region $A\cup B$. We referred to these surfaces as configuration-$1$. The HRT surfaces (shown as red curves) in this configuration is the union of two surfaces each of which go through the black brane from one asymptotic region to another. The shaded region denotes the entanglement wedge, which is disconnected in this configuration. }
    \label{fig-conf-1}
\end{figure}

\begin{figure}
    \centering
    \includegraphics[scale=.65]{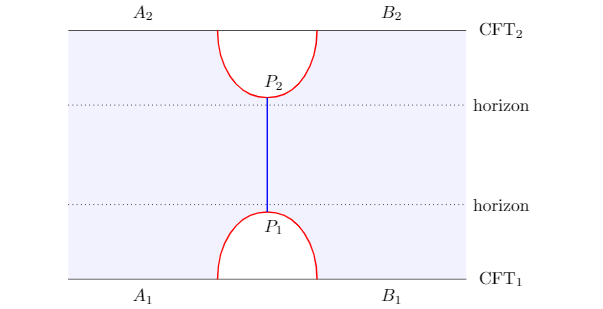}
    \caption{The pictorial representation of the second possibility of the HRT surfaces corresponding to region $A\cup B$. We referred to these surfaces as configuration-$2$. The HRT surfaces (shown as red curves) in this configuration is the union of two surfaces, both of which remain in the asymptotic regions. As a result, the entanglement wedge (shaded region) in this configuration is connected. The blue surface is the cross-section of the entanglement wedge.  }
    \label{fig-conf-2}
\end{figure}


One possible HRT surface is the union of two surfaces each of which go through the black brane. The pictorial representation of this configuration is shown in Fig.~(\ref{fig-conf-1}). The other possibility consists of two surfaces that are entirely in each of the exterior regions. This is shown in Fig.~(\ref{fig-conf-2}). 
The HRT surface at any instant of time is the configuration with the smallest area. The total area of surfaces in the two possible configurations is
\begin{align}
  \text{Configuration-$1$:} \quad\quad  &\text{Area} \, = \, \frac{4\pi}{\beta } \, \times \, 2t \, , \label{eq-conf-1}\\
   \text{Configuration-$2$:} \quad\quad  &\text{Area} \, = \, \frac{4\pi}{\beta } \, \times \, \ell \, . \label{eq-conf-2} 
\end{align}


Now using these results, we deduce that HRT surfaces are the surfaces in the configuration-$1$ for $t < \ell/2$ whereas the HRT surfaces are the surfaces in the configuration-$2$ for $t > \ell/2$. Note that the entanglement wedge in the configuration-$1$ is disconnected as shown in Fig.~(\ref{fig-conf-1}). As we discussed in Sec.~(\ref{sec-sr-ent-wedge}), the reflected entropy is zero if the entanglement wedge is disconnected. Therefore, the reflected entropy is only non-zero for $t > \ell/2$. In the next subsection, we explicitly calculate the entanglement wedge cross-section and reflected entropy for this range of time.



\subsection{Time dependence of the entanglement wedge cross-section}


The entanglement wedge cross-section that we are interested in is shown as a blue curve in Fig.~(\ref{fig-conf-2}) and its end-points are denoted by $P_{1}$ and $P_{2}$. Note that the point $P_{1}$ is the bulk turning point of the minimal area surface anchored on the boundary $1$ at points $(\rho \, , \, t \, , \, x) \, = \, (\infty \, , t \, ,  \, \ell/2 )$ and $(\rho \, , \, t \, , \, x) \, = \, (\infty \, , t \, ,  \, - \ell/2 )$. Owing to symmetry, the coordinates of the point $P_{1}$ are
\begin{align}
    P_{1} \, : \, (\rho \, , \, t \, , \, x) \, = \, (\rho_{*} \, , t \, ,  \, 0 ) \, ,
\end{align}
and it was found in \cite{Hubeny:2007xt} that $\rho_{*}$ is given by
\begin{align}
    \cosh\rho_{*} \, = \, \coth \left(\pi\ell/\beta\right) \, .\label{eq-rho-star}
\end{align}
Now using Eqs.~\eqref{eq-z-1}-\eqref{eq-x0-1}, we find that the point $P_{1}$ in Poincare coordinates is 
\begin{align}
    P_{1} \, : \, (z \, , \, x_{0} \, , \, x_{1}) \, = \, (1/\cosh\rho_{*} \, , \tanh\rho_{*} \,  \cosh\left(2\pi t/\beta\right) \, ,  \, \tanh\rho_{*} \,  \sinh\left(2\pi t/\beta\right) ) \, .
\end{align}
Similarly, the point $P_{2}$ in Poincare coordinates is
\begin{align}
    P_{2} \, : \, (z \, , \, x_{0} \, , \, x_{1}) \, = \, (1/\cosh\rho_{*} \, , \tanh\rho_{*} \,  \cosh\left(2\pi t/\beta\right) \, ,  \, -\tanh\rho_{*} \,  \sinh\left(2\pi t/\beta\right) ) \, .
\end{align}

In Poincare coordinates, the geodesic connecting $P_{1}$ and $P_{2}$ is a segment of a semi-circle at equal $x_{0}$-slice. That is, this geodesic satisfies
\begin{align}
    z \, = \, \sqrt{R^{2}-x_{1}^{2} \, } \, , \quad\quad\quad \text{and} \quad\quad\quad \, x_{0} \, = \, \tanh\rho_{*} \, \cosh\left(2\pi t/\beta\right) \, ,
\end{align}
where 
\begin{align}
    R^{2} \, = \, 1 + \tanh^{2}\rho_{*} \, \sinh^{2}\left(2\pi t/\beta\right) \, .\label{eq-Rad}
\end{align}
The length of this geodesic between points $P_{1}$ and $P_{2} \, $, computed using the metric in Eq.~\eqref{eq-met-poin}, is 
\begin{align}
    L_{12} \, = \, 2\, \log \left( \cosh\rho_{*} \, R \,  + \sqrt{\cosh^{2}\rho_{*} \, R^{2} \, - \, 1 \,} \right) \, .
\end{align}
In the scaling limit, this becomes
\begin{align}
    L_{12} \, = \, \frac{4\pi}{\beta} \left(t \, - \, \ell/2\right) \, .
\end{align}

Now using Eq.~\eqref{eq-ent-wc}, the entanglement wedge cross-section is then given
\begin{align}
    E_{W}(A|B) \, = \, \frac{\pi}{G\beta} \, \left(t-\ell/2\right) \, .
\end{align}
Combining the above result with the holographic formula for the reflected entropy, that is Eq.~\eqref{eq-ref-ent-holo}, and using the standard formula in AdS$_{3}$-CFT$_{2}$ correspondence,
\begin{align}
    c \, = \, \frac{3}{2G} \, ,
\end{align}
we get
\begin{align}
    S_{R}(A|B) \, = \, 4 \, s_{eq} \, \left(t-\ell/2\right) \, ,
\end{align}
where we have also used Eq.~\eqref{eq-seq}.

To summarize, we find that the time-dependent reflected entropy in the limit $L\to\infty$ and finite $\ell$ is given by 
\begin{align}
    S_{R}(A|B) \, = \, 4 s_{eq} \, \times \label{eq-sr-holo-fin}
\begin{cases}
\, 0 \quad&\text{for $\quad t <\frac{\ell}{2}$} \, ,\\[0ex]
\, t - \frac{\ell}{2} \quad&\text{for $\quad t > \frac{\ell}{2}$} \, . 
\end{cases}  
\end{align}

This finishes our discussion of the time dependence of the holographic reflected entropy when $A$ and $B$ are semi-infinite regions. We find that our result matches the result in Eq.~\eqref{eq-fin-case-1-gen}. This should not be surprising, because as we discussed in Sec.~(\ref{sec-case-1}), the time dependence of the reflected entropy for two semi-infinite regions is completely fixed by the conformal symmetry and should be same for all CFTs. 

\section{Discussion} \label{sec-disc}

In this paper, we have studied the time dependence of the reflected entropy in a thermal double model of \cite{hartman-maldacena}. We have focused on $(1+1)$-dimensional rational theories and holographic theories. For rational CFTs, we used the replica trick to calculate the reflected entropy. We found that the time dependence of the reflected entropy is the same as that of the mutual information. For holographic theories, we used the relation between the reflected entropy and the entanglement wedge cross-section to calculate the reflected entropy. As a future direction, it would be interesting to study the time evolution of the reflected entropy in CFTs which are neither rational nor holographic. 

There are many possible directions in which our work can be extended. For example, it has been argued that the dynamics of the entanglement entropy for holographic states can be described in terms of a minimal membrane \cite{Mezei-1,Mezei-3}. Since the reflected entropy in the holographic theories is also given by an extremization process, it is fair to expect that a similar membrane description holds for the dynamics of the reflected entropy\footnote{We thank M. Mezei for a discussion about it.}. It will be interesting to understand this membrane description in detail.

\textit{Note:} Similar calculations of the time dependence of the reflected entropy are performed in an independent work \cite{Kudler-Flam:2020url} that appeared on arXiv simultaneously with the first version of this paper. 


\vskip .3cm
{\bf Acknowledgments} 
It is a pleasure to thank C. Akers, N. Bao, T. Hartman, and P. Rath for helpful discussions, and to T. Hartman for useful feedback on a draft of this manuscript. This work was supported by the US Department of Energy under grant number DE-SC$0014123$.

\bibliographystyle{utcaps}
\bibliography{all}

\providecommand{\href}[2]{#2}\begingroup\raggedright\begin{thebibliography}{10}

\bibitem{calabrese-cardy}
P.~Calabrese and J.~L. Cardy, ``{Evolution of entanglement entropy in
  one-dimensional systems},''
  \href{http://dx.doi.org/10.1088/1742-5468/2005/04/P04010}{{\em J. Stat.
  Mech.} {\bf 0504} (2005)  P04010},
\href{http://arxiv.org/abs/cond-mat/0503393}{{\tt arXiv:cond-mat/0503393
  [cond-mat]}}.

\bibitem{gq-2}
P.~Calabrese and J.~Cardy, ``{Entanglement entropy and conformal field
  theory},'' \href{http://dx.doi.org/10.1088/1751-8113/42/50/504005}{{\em J.
  Phys.} {\bf A42} (2009)  504005},
\href{http://arxiv.org/abs/0905.4013}{{\tt arXiv:0905.4013
  [cond-mat.stat-mech]}}.

\bibitem{Holzhey:1994we}
C.~Holzhey, F.~Larsen, and F.~Wilczek, ``{Geometric and renormalized entropy in
  conformal field theory},''
  \href{http://dx.doi.org/10.1016/0550-3213(94)90402-2}{{\em Nucl. Phys. B}
  {\bf 424} (1994)  443--467}, \href{http://arxiv.org/abs/hep-th/9403108}{{\tt
  arXiv:hep-th/9403108}}.

\bibitem{hartman-maldacena}
T.~Hartman and J.~Maldacena, ``{Time Evolution of Entanglement Entropy from
  Black Hole Interiors},''
  \href{http://dx.doi.org/10.1007/JHEP05(2013)014}{{\em JHEP} {\bf 05} (2013)
  014},
\href{http://arxiv.org/abs/1303.1080}{{\tt arXiv:1303.1080 [hep-th]}}.

\bibitem{gq-11}
C.~T. Asplund and A.~Bernamonti, ``{Mutual information after a local quench in
  conformal field theory},''
  \href{http://dx.doi.org/10.1103/PhysRevD.89.066015}{{\em Phys. Rev.} {\bf
  D89} (2014) no.~6, 066015},
\href{http://arxiv.org/abs/1311.4173}{{\tt arXiv:1311.4173 [hep-th]}}.

\bibitem{gq-12}
S.~Leichenauer and M.~Moosa, ``{Entanglement Tsunami in (1+1)-Dimensions},''
  \href{http://dx.doi.org/10.1103/PhysRevD.92.126004}{{\em Phys. Rev.} {\bf
  D92} (2015)  126004},
\href{http://arxiv.org/abs/1505.04225}{{\tt arXiv:1505.04225 [hep-th]}}.

\bibitem{hartman-2}
C.~T. Asplund, A.~Bernamonti, F.~Galli, and T.~Hartman, ``{Entanglement
  Scrambling in 2d Conformal Field Theory},''
  \href{http://dx.doi.org/10.1007/JHEP09(2015)110}{{\em JHEP} {\bf 09} (2015)
  110},
\href{http://arxiv.org/abs/1506.03772}{{\tt arXiv:1506.03772 [hep-th]}}.

\bibitem{gq-3}
J.~Abajo-Arrastia, J.~Aparicio, and E.~Lopez, ``{Holographic Evolution of
  Entanglement Entropy},''
  \href{http://dx.doi.org/10.1007/JHEP11(2010)149}{{\em JHEP} {\bf 11} (2010)
  149},
\href{http://arxiv.org/abs/1006.4090}{{\tt arXiv:1006.4090 [hep-th]}}.

\bibitem{Albash:2010mv}
T.~Albash and C.~V. Johnson, ``{Evolution of Holographic Entanglement Entropy
  after Thermal and Electromagnetic Quenches},''
  \href{http://dx.doi.org/10.1088/1367-2630/13/4/045017}{{\em New J. Phys.}
  {\bf 13} (2011)  045017},
\href{http://arxiv.org/abs/1008.3027}{{\tt arXiv:1008.3027 [hep-th]}}.

\bibitem{gq-4}
V.~Balasubramanian, A.~Bernamonti, J.~de~Boer, N.~Copland, B.~Craps,
  E.~Keski-Vakkuri, B.~Muller, A.~Schafer, M.~Shigemori, and W.~Staessens,
  ``{Thermalization of Strongly Coupled Field Theories},''
  \href{http://dx.doi.org/10.1103/PhysRevLett.106.191601}{{\em Phys. Rev.
  Lett.} {\bf 106} (2011)  191601},
\href{http://arxiv.org/abs/1012.4753}{{\tt arXiv:1012.4753 [hep-th]}}.

\bibitem{gq-5}
V.~Balasubramanian, A.~Bernamonti, J.~de~Boer, N.~Copland, B.~Craps,
  E.~Keski-Vakkuri, B.~Muller, A.~Schafer, M.~Shigemori, and W.~Staessens,
  ``{Holographic Thermalization},''
  \href{http://dx.doi.org/10.1103/PhysRevD.84.026010}{{\em Phys. Rev.} {\bf
  D84} (2011)  026010},
\href{http://arxiv.org/abs/1103.2683}{{\tt arXiv:1103.2683 [hep-th]}}.

\bibitem{gq-6}
V.~Balasubramanian, A.~Bernamonti, N.~Copland, B.~Craps, and F.~Galli,
  ``{Thermalization of mutual and tripartite information in strongly coupled
  two dimensional conformal field theories},''
  \href{http://dx.doi.org/10.1103/PhysRevD.84.105017}{{\em Phys. Rev.} {\bf
  D84} (2011)  105017},
\href{http://arxiv.org/abs/1110.0488}{{\tt arXiv:1110.0488 [hep-th]}}.

\bibitem{gq-7}
A.~Allais and E.~Tonni, ``{Holographic evolution of the mutual information},''
  \href{http://dx.doi.org/10.1007/JHEP01(2012)102}{{\em JHEP} {\bf 01} (2012)
  102},
\href{http://arxiv.org/abs/1110.1607}{{\tt arXiv:1110.1607 [hep-th]}}.

\bibitem{gq-9}
H.~Liu and S.~J. Suh, ``{Entanglement Tsunami: Universal Scaling in Holographic
  Thermalization},''
  \href{http://dx.doi.org/10.1103/PhysRevLett.112.011601}{{\em Phys. Rev.
  Lett.} {\bf 112} (2014)  011601},
\href{http://arxiv.org/abs/1305.7244}{{\tt arXiv:1305.7244 [hep-th]}}.

\bibitem{gq-10}
H.~Liu and S.~J. Suh, ``{Entanglement growth during thermalization in
  holographic systems},''
  \href{http://dx.doi.org/10.1103/PhysRevD.89.066012}{{\em Phys. Rev.} {\bf
  D89} (2014) no.~6, 066012},
\href{http://arxiv.org/abs/1311.1200}{{\tt arXiv:1311.1200 [hep-th]}}.

\bibitem{gq-13}
V.~Ziogas, ``{Holographic mutual information in global Vaidya-BTZ spacetime},''
  \href{http://dx.doi.org/10.1007/JHEP09(2015)114}{{\em JHEP} {\bf 09} (2015)
  114},
\href{http://arxiv.org/abs/1507.00306}{{\tt arXiv:1507.00306 [hep-th]}}.

\bibitem{gq-14}
M.~Rangamani, M.~Rozali, and A.~Vincart-Emard, ``{Dynamics of Holographic
  Entanglement Entropy Following a Local Quench},''
  \href{http://dx.doi.org/10.1007/JHEP04(2016)069}{{\em JHEP} {\bf 04} (2016)
  069},
\href{http://arxiv.org/abs/1512.03478}{{\tt arXiv:1512.03478 [hep-th]}}.

\bibitem{gq-15}
M.~R. Tanhayi, ``{Thermalization of Mutual Information in Hyperscaling
  Violating Backgrounds},''
  \href{http://dx.doi.org/10.1007/JHEP03(2016)202}{{\em JHEP} {\bf 03} (2016)
  202},
\href{http://arxiv.org/abs/1512.04104}{{\tt arXiv:1512.04104 [hep-th]}}.

\bibitem{gq-16}
S.~Leichenauer, M.~Moosa, and M.~Smolkin, ``{Dynamics of the Area Law of
  Entanglement Entropy},''
  \href{http://dx.doi.org/10.1007/JHEP09(2016)035}{{\em JHEP} {\bf 09} (2016)
  035},
\href{http://arxiv.org/abs/1604.00388}{{\tt arXiv:1604.00388 [hep-th]}}.

\bibitem{gq-17}
A.~Sivaramakrishnan, ``{Localized Excitations from Localized Unitary
  Operators},'' \href{http://dx.doi.org/10.1016/j.aop.2017.03.012}{{\em Annals
  Phys.} {\bf 381} (2017)  41--67},
\href{http://arxiv.org/abs/1604.00965}{{\tt arXiv:1604.00965 [hep-th]}}.

\bibitem{gq-18}
M.~Mezei and D.~Stanford, ``{On entanglement spreading in chaotic systems},''
  \href{http://dx.doi.org/10.1007/JHEP05(2017)065}{{\em JHEP} {\bf 05} (2017)
  065},
\href{http://arxiv.org/abs/1608.05101}{{\tt arXiv:1608.05101 [hep-th]}}.

\bibitem{gq-20}
S.~F. Lokhande, G.~W.~J. Oling, and J.~F. Pedraza, ``{Linear response of
  entanglement entropy from holography},''
  \href{http://dx.doi.org/10.1007/JHEP10(2017)104}{{\em JHEP} {\bf 10} (2017)
  104},
\href{http://arxiv.org/abs/1705.10324}{{\tt arXiv:1705.10324 [hep-th]}}.

\bibitem{gq-22}
M.~Flory, J.~Erdmenger, D.~Fernandez, E.~Megias, A.-K. Straub, and
  P.~Witkowski, ``{Time dependence of entanglement for steady state formation
  in AdS$_3$/CFT$_2$},'' in {\em {3rd Karl Schwarzschild Meeting on
  Gravitational Physics and the Gauge/Gravity Correspondence (KSM 2017)
  Frankfurt am Main, Germany, July 24-28, 2017}}.
\newblock 2017.
\newblock \href{http://arxiv.org/abs/1709.08614}{{\tt arXiv:1709.08614
  [hep-th]}}.
\newblock
\url{http://inspirehep.net/record/1625285/files/arXiv:1709.08614.pdf}.
\newblock

\bibitem{Mezei-1}
M.~Mezei, ``{Membrane theory of entanglement dynamics from holography},''
  \href{http://dx.doi.org/10.1103/PhysRevD.98.106025}{{\em Phys. Rev.} {\bf
  D98} (2018) no.~10, 106025},
\href{http://arxiv.org/abs/1803.10244}{{\tt arXiv:1803.10244 [hep-th]}}.

\bibitem{Mezei-3}
M.~Mezei and J.~Virrueta, ``{Exploring the Membrane Theory of Entanglement
  Dynamics},''
\href{http://arxiv.org/abs/1912.11024}{{\tt arXiv:1912.11024 [hep-th]}}.

\bibitem{10.21468/SciPostPhys.4.3.017}
V.~Alba and P.~Calabrese, ``{Entanglement dynamics after quantum quenches in
  generic integrable systems},''
  \href{http://dx.doi.org/10.21468/SciPostPhys.4.3.017}{{\em SciPost Phys.}
  {\bf 4} (2018)  17}. \url{https://scipost.org/10.21468/SciPostPhys.4.3.017}.

\bibitem{Vidal:2002zz}
G.~Vidal and R.~Werner, ``{Computable measure of entanglement},''
  \href{http://dx.doi.org/10.1103/PhysRevA.65.032314}{{\em Phys. Rev. A} {\bf
  65} (2002)  032314}, \href{http://arxiv.org/abs/quant-ph/0102117}{{\tt
  arXiv:quant-ph/0102117}}.

\bibitem{Coser:2014gsa}
A.~Coser, E.~Tonni, and P.~Calabrese, ``{Entanglement negativity after a global
  quantum quench},''
  \href{http://dx.doi.org/10.1088/1742-5468/2014/12/P12017}{{\em J. Stat.
  Mech.} {\bf 1412} (2014) no.~12, P12017},
\href{http://arxiv.org/abs/1410.0900}{{\tt arXiv:1410.0900
  [cond-mat.stat-mech]}}.

\bibitem{faulkner}
S.~Dutta and T.~Faulkner, ``{A canonical purification for the entanglement
  wedge cross-section},''
\href{http://arxiv.org/abs/1905.00577}{{\tt arXiv:1905.00577 [hep-th]}}.

\bibitem{sr-local-1}
Y.~Kusuki and K.~Tamaoka, ``{Dynamics of Entanglement Wedge Cross Section from
  Conformal Field Theories},''
\href{http://arxiv.org/abs/1907.06646}{{\tt arXiv:1907.06646 [hep-th]}}.

\bibitem{sr-local-2}
Y.~Kusuki and K.~Tamaoka, ``{Entanglement Wedge Cross Section from CFT:
  Dynamics of Local Operator Quench},''
\href{http://arxiv.org/abs/1909.06790}{{\tt arXiv:1909.06790 [hep-th]}}.

\bibitem{Akers:2019gcv}
C.~Akers and P.~Rath, ``{Entanglement Wedge Cross Sections Require Tripartite
  Entanglement},''
\href{http://arxiv.org/abs/1911.07852}{{\tt arXiv:1911.07852 [hep-th]}}.

\bibitem{Calabrese:2009qy}
P.~Calabrese and J.~Cardy, ``{Entanglement entropy and conformal field
  theory},'' \href{http://dx.doi.org/10.1088/1751-8113/42/50/504005}{{\em J.
  Phys.} {\bf A42} (2009)  504005},
\href{http://arxiv.org/abs/0905.4013}{{\tt arXiv:0905.4013
  [cond-mat.stat-mech]}}.

\bibitem{eop-1}
T.~Takayanagi and K.~Umemoto, ``{Entanglement of purification through
  holographic duality},''
  \href{http://dx.doi.org/10.1038/s41567-018-0075-2}{{\em Nature Phys.} {\bf
  14} (2018) no.~6, 573--577},
\href{http://arxiv.org/abs/1708.09393}{{\tt arXiv:1708.09393 [hep-th]}}.

\bibitem{eop-2}
P.~Nguyen, T.~Devakul, M.~G. Halbasch, M.~P. Zaletel, and B.~Swingle,
  ``{Entanglement of purification: from spin chains to holography},''
  \href{http://dx.doi.org/10.1007/JHEP01(2018)098}{{\em JHEP} {\bf 01} (2018)
  098},
\href{http://arxiv.org/abs/1709.07424}{{\tt arXiv:1709.07424 [hep-th]}}.

\bibitem{Maldacena:2001kr}
J.~M. Maldacena, ``{Eternal black holes in anti-de Sitter},''
  \href{http://dx.doi.org/10.1088/1126-6708/2003/04/021}{{\em JHEP} {\bf 04}
  (2003)  021},
\href{http://arxiv.org/abs/hep-th/0106112}{{\tt arXiv:hep-th/0106112
  [hep-th]}}.

\bibitem{Banados:1992wn}
M.~Banados, C.~Teitelboim, and J.~Zanelli, ``{The Black hole in
  three-dimensional space-time},''
  \href{http://dx.doi.org/10.1103/PhysRevLett.69.1849}{{\em Phys. Rev. Lett.}
  {\bf 69} (1992)  1849--1851},
\href{http://arxiv.org/abs/hep-th/9204099}{{\tt arXiv:hep-th/9204099
  [hep-th]}}.

\bibitem{Maldacena:1998bw}
J.~M. Maldacena and A.~Strominger, ``{AdS(3) black holes and a stringy
  exclusion principle},''
  \href{http://dx.doi.org/10.1088/1126-6708/1998/12/005}{{\em JHEP} {\bf 12}
  (1998)  005},
\href{http://arxiv.org/abs/hep-th/9804085}{{\tt arXiv:hep-th/9804085
  [hep-th]}}.

\bibitem{Parikh:2012kg}
M.~Parikh and P.~Samantray, ``{Rindler-AdS/CFT},''
  \href{http://dx.doi.org/10.1007/JHEP10(2018)129}{{\em JHEP} {\bf 10} (2018)
  129},
\href{http://arxiv.org/abs/1211.7370}{{\tt arXiv:1211.7370 [hep-th]}}.

\bibitem{Hubeny:2007xt}
V.~E. Hubeny, M.~Rangamani, and T.~Takayanagi, ``{A Covariant holographic
  entanglement entropy proposal},''
  \href{http://dx.doi.org/10.1088/1126-6708/2007/07/062}{{\em JHEP} {\bf 07}
  (2007)  062},
\href{http://arxiv.org/abs/0705.0016}{{\tt arXiv:0705.0016 [hep-th]}}.

\bibitem{Kudler-Flam:2020url}
J.~Kudler-Flam, Y.~Kusuki, and S.~Ryu, ``{Correlation measures and the
  entanglement wedge cross-section after quantum quenches in two-dimensional
  conformal field theories},''
\href{http://arxiv.org/abs/2001.05501}{{\tt arXiv:2001.05501 [hep-th]}}.

\end{thebibliography}\endgroup
\end{document}